\documentclass[12pt]{article}
\usepackage{amsmath}
\usepackage{latexsym}
\usepackage{amsfonts}
\usepackage[normalem]{ulem}
\usepackage{array}
\usepackage{amssymb}
\usepackage{graphicx}
\usepackage[backend=biber,
style=numeric,
sorting=none,
isbn=false,
doi=false,
url=false,
]{biblatex}

\usepackage{subfig}
\usepackage{wrapfig}
\usepackage{wasysym}
\usepackage{enumitem}
\usepackage{adjustbox}
\usepackage{ragged2e}
\usepackage[svgnames,table]{xcolor}
\usepackage{tikz}
\usepackage{longtable}
\usepackage{changepage}
\usepackage{setspace}
\usepackage{hhline}
\usepackage{multicol}
\usepackage{tabto}
\usepackage{float}
\usepackage{multirow}
\usepackage{makecell}
\usepackage{fancyhdr}
\usepackage[toc,page]{appendix}
\usepackage[hidelinks]{hyperref}
\usetikzlibrary{shapes.symbols,shapes.geometric,shadows,arrows.meta}
\tikzset{>={Latex[width=1.5mm,length=2mm]}}
\usepackage{flowchart}\usepackage[paperheight=11.0in,paperwidth=8.5in,left=1.0in,right=1.0in,top=1.0in,bottom=1.0in,headheight=1in]{geometry}
\usepackage[utf8]{inputenc}
\usepackage[T1]{fontenc}
\TabPositions{0.5in,1.0in,1.5in,2.0in,2.5in,3.0in,3.5in,4.0in,4.5in,5.0in,5.5in,6.0in,}

\setlistdepth{9}
\renewlist{enumerate}{enumerate}{9}
		\setlist[enumerate,1]{label=\arabic*)}
		\setlist[enumerate,2]{label=\alph*)}
		\setlist[enumerate,3]{label=(\roman*)}
		\setlist[enumerate,4]{label=(\arabic*)}
		\setlist[enumerate,5]{label=(\Alph*)}
		\setlist[enumerate,6]{label=(\Roman*)}
		\setlist[enumerate,7]{label=\arabic*}
		\setlist[enumerate,8]{label=\alph*}
		\setlist[enumerate,9]{label=\roman*}

\renewlist{itemize}{itemize}{9}
		\setlist[itemize]{label=$\cdot$}
		\setlist[itemize,1]{label=\textbullet}
		\setlist[itemize,2]{label=$\circ$}
		\setlist[itemize,3]{label=$\ast$}
		\setlist[itemize,4]{label=$\dagger$}
		\setlist[itemize,5]{label=$\triangleright$}
		\setlist[itemize,6]{label=$\bigstar$}
		\setlist[itemize,7]{label=$\blacklozenge$}
		\setlist[itemize,8]{label=$\prime$}

\begin{document}
{\fontsize{16pt}{19.2pt}\selectfont \textcolor[HTML]{4472C4}{Title: Ripple oscillations in the left temporal neocortex are associated with impaired verbal episodic memory encoding}\par}\par

\vspace{\baselineskip}
{\fontsize{11pt}{13.2pt}\selectfont \textcolor[HTML]{1B2432}{Zachary J. Waldman\textsuperscript{1}, Liliana Camarillo-Rodriguez\textsuperscript{1}, Inna Chervenova\textsuperscript{2}, Brent Berry\textsuperscript{6,7}, Shoichi Shimamoto\textsuperscript{1}, Bahareh Elahian\textsuperscript{1}, Michal Kucewicz\textsuperscript{6,7}, Chaitanya Ganne\textsuperscript{3}, Xiao-Song He\textsuperscript{3}, Leon A. Davis\textsuperscript{8}, Joel Stein\textsuperscript{9}, Sandhitsu Das\textsuperscript{10,11}, Richard Gorniak\textsuperscript{4}, Ashwini D. Sharan\textsuperscript{5}, Robert Gross\textsuperscript{13}, Cory S. Inman\textsuperscript{13}, Bradley C. Lega\textsuperscript{14}, Kareem Zaghloul\textsuperscript{15}, Barbara C. Jobst\textsuperscript{16}, Katheryn A. Davis\textsuperscript{12}, Paul Wanda\textsuperscript{8}, Mehraneh Khadjevand\textsuperscript{6,7}, Joseph Tracy\textsuperscript{3}, Daniel S. Rizzuto\textsuperscript{8}, Gregory Worrell\textsuperscript{6,7}, Michael Sperling\textsuperscript{3}, Shennan A. Weiss\textsuperscript{1}}\par}\par

\vspace{\baselineskip}
{\fontsize{11pt}{13.2pt}\selectfont 1. Dept. of Neurology and Neuroscience, 2. Dept. of Pharmacology $\&$  Experimental Therapeutics, 3. Dept. of Neurology, 4. Dept. of Radiology, 5. Dept. of Neurosurgery, Thomas Jefferson University, Philadelphia, PA USA 19107. 6. Dept. of Neurology, Mayo Systems Electrophysiology Laboratory (MSEL). 7. Dept. of Physiology and Biomedical Engineering, Mayo Clinic, Rochester, MN USA 55905. 8. Dept. of Psychology, 9. Department of Radiology, 10. Penn Image Computing and Science Laboratory, Department of Radiology, 11. Penn Memory Center, Department of Neurology, 12. Dept. of Neurology, University of Pennsylvania, Philadelphia, PA USA 19104. 13. Emory University, Dept. of Neurosurgery, Atlanta, GA USA 30322. 14. University of Texas Southwestern Medical Center, Dept. of Neurosurgery, Dallas, TX USA 75390. 15. Surgical Neurology Branch, NINDS, NIH, Bethesda, MD USA 20892. 16. Dartmouth-Hitchcock Medical Center, Dept. of Neurology, Lebanon, NH USA 03756. \par}\par

\vspace{\baselineskip}
{\fontsize{11pt}{13.2pt}\selectfont Corresponding Author: \par}\par

{\fontsize{11pt}{13.2pt}\selectfont Dr. Shennan A. Weiss\par}\par

{\fontsize{11pt}{13.2pt}\selectfont 901 Walnut Street, PA 19107\par}\par

{\fontsize{11pt}{13.2pt}\selectfont Suite 400\par}\par

{\fontsize{11pt}{13.2pt}\selectfont Phone: (215) 503-7960\par}\par

{\fontsize{11pt}{13.2pt}\selectfont Email: \href{mailto:Shennan.Weiss@jefferson.edu}{Shennan.Weiss@jefferson.edu}\par}\par

\vspace{\baselineskip}
{\fontsize{11pt}{13.2pt}\selectfont Abbreviated title: Intracranial EEG ripples and disrupted memory encoding\par}\par

\vspace{\baselineskip}
{\fontsize{11pt}{13.2pt}\selectfont Key Words: verbal memory, epilepsy, high-frequency oscillation, epileptiform discharge \par}\par

\vspace{\baselineskip}
{\fontsize{11pt}{13.2pt}\selectfont \textcolor[HTML]{FF0000}{Number of words: Introduction (411/600 words), Materials and methods (1544/1200):700 words), Results (997/1200 words), Discussion (1176 words). Total: 4128/4000 words.}\par}\par

\vspace{\baselineskip}

\vspace{\baselineskip}

\vspace{\baselineskip}

 %%%%%%%%%%%%  Starting New Page here %%%%%%%%%%%%%%

\newpage

\vspace{\baselineskip}{\fontsize{11pt}{13.2pt}\selectfont \textbf{Summary and Key Words (299/300 words)}\par}\par

{\fontsize{11pt}{13.2pt}\selectfont Background:\textcolor[HTML]{1B2432}{ We sought to determine if ripple oscillations (80-120Hz), detected in intracranial EEG (iEEG) recordings of epilepsy patients, correlate with an enhancement or disruption of verbal episodic memory encoding.}\par}\par

\vspace{\baselineskip}
{\fontsize{11pt}{13.2pt}\selectfont Methods:\textcolor[HTML]{1B2432}{ We defined ripple and spike events in depth iEEG recordings during list learning in 107 patients with focal epilepsy.  We used logistic regression models (LRMs) to investigate the relationship between the occurrence of ripple and spike events during word presentation and the odds of successful word recall following a distractor epoch, and included the seizure onset zone (SOZ) as a covariate in the LRMs.}\par}\par

\vspace{\baselineskip}
{\fontsize{11pt}{13.2pt}\selectfont Results:\textcolor[HTML]{1B2432}{ We detected events during 58,312 word presentation trials from 7,630 unique electrode sites. The probability of ripple on spike (RonS) events was increased in the seizure onset zone (SOZ, p<0.04). In the left temporal neocortex RonS events during word presentation corresponded with a decrease in the odds ratio (OR) of successful recall, however this effect only met significance in the SOZ (OR of word recall 0.71, 95$\%$  CI: 0.59-0.85, n=158 events, adaptive Hochberg} \textcolor[HTML]{1B2432}{p<0.01). Ripple on oscillation events (RonO) that occurred in the left temporal neocortex non-SOZ also correlated with decreased odds of successful recall (OR 0.52, 95$\%$  CI: 0.34-0.80, n=140, }adaptive Hochberg ,\textcolor[HTML]{1B2432}{ p<0.01). Spikes and RonS that occurred during word presentation in the left middle temporal gyrus during word presentation correlated with the most significant decrease in the odds of successful recall, irrespective of the location of the SOZ (}adaptive Hochberg, \textcolor[HTML]{1B2432}{p<0.01). }\par}\par

\vspace{\baselineskip}
{\fontsize{11pt}{13.2pt}\selectfont Conclusion: \textcolor[HTML]{1B2432}{Ripples and spikes generated in left temporal neocortex are associated with impaired verbal episodic memory encoding. Although physiological and pathological ripple oscillations were not distinguished during cognitive tasks, our results show an association of undifferentiated ripples with impaired encoding. The effect was sometimes specific to regions outside the SOZ, suggesting that widespread effects of epilepsy outside the SOZ may contribute to cognitive impairment.}\par}\par

\vspace{\baselineskip}
{\fontsize{10pt}{12.0pt}\selectfont \textbf{Key Words: }{\fontsize{11pt}{13.2pt}\selectfont verbal memory, epilepsy, high-frequency oscillation, epileptiform discharge\par}\par}\par

\vspace{\baselineskip}

\vspace{\baselineskip}

\vspace{\baselineskip}

\vspace{\baselineskip}

\vspace{\baselineskip}

\vspace{\baselineskip}

\vspace{\baselineskip}

\vspace{\baselineskip}

\vspace{\baselineskip}

\vspace{\baselineskip}

\vspace{\baselineskip}

\vspace{\baselineskip}

\vspace{\baselineskip}

\vspace{\baselineskip}

\vspace{\baselineskip}

\vspace{\baselineskip}

\vspace{\baselineskip}

\vspace{\baselineskip}

\vspace{\baselineskip}

\vspace{\baselineskip}

\vspace{\baselineskip}
\setstretch{2.0}
{\fontsize{11pt}{13.2pt}\selectfont \textbf{Introduction}\par}\par

\tab {\fontsize{11pt}{13.2pt}\selectfont Brief (1-50 msec) neurophysiological events such as action potentials in single neurons\textsuperscript{1} or high frequency oscillations (80-600 Hz)\textsuperscript{2} can serve as a marker of memory formation or cognition. Conversely, other brief neurophysiological events, such as interictal epileptiform spikes, can act as markers of cognitive disruption. Patients with absence seizures, which consist of long trains of generalized spikes, experience inattention and a disruption of consciousness\textsuperscript{3}. Focal interictal spikes are not typically perceived and do not interfere with consciousness, but may interfere with cognition\textsuperscript{4-6}. For example, human studies of verbal episodic memory have reported correlations between interictal discharge rate increases during memory encoding and impaired word recall\textsuperscript{6}. It is unknown if other brief neurophysiological events, besides epileptiform spikes, can also serve as markers of disrupted cognition. \par}\par

\tab {\fontsize{11pt}{13.2pt}\selectfont In experimental animals, ripple oscillations (80-200 Hz), when generated in area CA1 of the hippocampus during the sharp wave ripple complex (SpW-R), are known for mediating memory encoding\textsuperscript{7,8}, consolidation\textsuperscript{9}, and recall\textsuperscript{10}. In patients with epilepsy, ripple oscillations detected in intracranial EEG (iEEG) recordings from unaffected hippocampus during sleep can also correlate with successful memory consolidation\textsuperscript{11}. Furthermore, during wakefulness, ripple oscillations have been shown to mediate cognitive processing and memory throughout the brain due to fast network synchronization\textsuperscript{12-14}. However, ripple oscillations also occur more often in brain regions known to be epileptogenic\textsuperscript{15}, regardlesss of whether they occur superimposed on interictal spikes (RonS), or the ripples are sumerimposed on background EEG oscillations (RonO)\textsuperscript{16}. \par}\par

{\fontsize{11pt}{13.2pt}\selectfont \tab As ripples can both subserve memory and demarcate epileptogenic tissue, the overall cognitive significance of the biomarker for patients with epilepsy is unclear. Limited evidence suggests that ripple rates do not positively correlate with memory function, but rather memory performance negatively correlates with ripple rates measured outside the seizure onset zone\textsuperscript{17}. Thus, ripple events may sometimes promote or possibly disrupt memory.\par}\par

{\fontsize{11pt}{13.2pt}\selectfont \tab In this study, we investigate a possible correlation between ripple and spike occurrences in macroelectrode iEEG recordings during word encoding trials of a list learning task, and changes in the probability of recall. No \textit{a priori} assumptions were made regarding the pathological or physiological nature of each ripple event. We utilized a large cohort of medically refractory focal epilepsy patients who performed a list learning free recall task while undergoing iEEG evaluation for epilepsy surgery with depth electrodes\textsuperscript{ 6,17}. We focused on delineating the neuroanatomical regions susceptible to modulation by ripple and spikes to better understand the precise effect of ripples, and to help resolve the functional hubs of the network that mediate human verbal episodic memory. \par}\par

\vspace{\baselineskip}
{\fontsize{11pt}{13.2pt}\selectfont \textbf{Materials and methods }\par}\par

{\fontsize{11pt}{13.2pt}\selectfont \textit{Subjects}\par}\par

{\fontsize{11pt}{13.2pt}\selectfont Patients undergoing iEEG monitoring as part of the pre-surgical treatment for drug-resistant epilepsy were recruited in this multi-center study. Data were collected from: Thomas Jefferson University Hospital (Philadelphia, PA), Mayo Clinic (Rochester, MN), Hospital of the University of Pennsylvania (Philadelphia, PA), Dartmouth-Hitchcock Medical Center (Lebanon, NH), Emory University Hospital (Atlanta, GA), University of Texas Southwestern Medical Center (Dallas, TX), and Columbia University Medical Center (New York, NY). The research protocol was approved by each respective IRB and informed consent was obtained from each subject. Electrophysiological recordings were collected from clinical subdural and depth electrodes (AdTech Inc., PMT Inc.).\par}\par

\vspace{\baselineskip}
{\fontsize{11pt}{13.2pt}\selectfont \textit{Electrode localization and defining the location of the SOZ}\par}\par

{\fontsize{11pt}{13.2pt}\selectfont Coordinates of surface and depth electrode contacts were obtained for all subjects from post-implantation CT scans. Pre-implantation volumetric T1-weighted MRI scans were co-registered to the CT scans as well as to the MNI152 standard brain to enable comparison of recording sites in a common space across subjects. Anatomic locations of the recording sites were derived by converting MNI coordinates to Talairach coordinates and querying the Talairach daemon. The seizure onset zone (SOZ) was clinically defined by visual inspection of ictal iEEG by clinicians at each of the data collection sites. \par}\par

\vspace{\baselineskip}
{\fontsize{11pt}{13.2pt}\selectfont \textit{Memory task}\par}\par

{\fontsize{11pt}{13.2pt}\selectfont Subjects participated in list learning delayed free recall memory tasks. In order to facilitate the comfort of the patient, the task was organized into blocks. During each block 12 randomly chosen words were displayed on a computer screen. Each word was displayed for 1600 msec, and the inter-word interval was jittered between 750-1000 msec. Lists were chosen from a pool of high-frequency nouns (available at http://memory.psych.upenn.edu/WordPools). Following the word display block the subjects performed an arithmetic distractor task for an average of 20 seconds. Participants were then provided with an average of 30 seconds to verbally recall the words in any order. Patients performed up to 25 blocks per session and some patients performed more than one session\textsuperscript{18}. \par}\par

{\fontsize{11pt}{13.2pt}\selectfont  \par}\par

{\fontsize{11pt}{13.2pt}\selectfont \textit{Intracranial EEG data}\par}\par

{\fontsize{11pt}{13.2pt}\selectfont Intracranial data were recorded using either a Nihon Kohden EEG-1200, Natus XLTek EMU 128 or Grass Aura-LTM64. The iEEG signals were sampled at either 500, 1000 or 1600 Hz and were referenced to a common contact placed either intracranially, on the scalp, or the mastoid process. A bipolar montage was calculated after recordings for each subject. All bipolar derivations with sampling frequency >500 Hz were low-pass filtered < 250 Hz and then downsampled to 500 Hz. To examine the effects of ripples and spikes on encoding, we analyzed iEEG during both the 750 msec of inter-word-interval, and 1600 msec during word display. We also analyzed iEEG during the entirety of the distractor and recall tasks for each block.\par}\par

\vspace{\baselineskip}
{\fontsize{11pt}{13.2pt}\selectfont \textit{Detecting candidate ripple events in iEEG recordings}\par}\par

{\fontsize{11pt}{13.2pt}\selectfont Prior to data preprocessing we first determined which depth electrode iEEG recordings exhibited a signal to noise ratio sufficient for accurate ripple detection (Supplementary Methods). \par}\par

{\fontsize{11pt}{13.2pt}\selectfont \tab Following the exclusion of recordings from electrode contacts with excessive high frequency artifacts we selected the time intervals of the recordings that would be subject to our analysis. During encoding, the jitter between each word trial varied between 750-1000 msec. Due to this variability we selected for our analysis of the word encoding trial iEEG recordings which concatenated the final 750 msec of pre-word interval with the 1600 msec of word presentation. For the distractor and recall epochs we divided the entire recording period into 2000 msec contiguous trials. \par}\par

{\fontsize{11pt}{13.2pt}\selectfont \tab We next generated a time-frequency (TF) plot of each trial using wavelet convolution in the time domain\textsuperscript{1 }(Supplementary Methods). To identify the candidate ripple events (ripple on spike [RonS] and ripple on oscillation [RonO]) in the time-frequency plot we used a power magnitude threshold of 1$\ast$ 10\textsuperscript{7} arbitrary units. This power threshold corresponds to a ripple with a peak amplitude of approximately 8 V\textsuperscript{20}. We chose this threshold to define only those events with good inter-rater reliability upon visual inspection\textsuperscript{15}. Detected ripple on oscillation events were inspected in Matlab™ (Natick, MA) by ZJW, and removed from the dataset if deemed artefactual. The sensitivity and precision of the resulting ripple and spike detections was independently validated by two coauthors SW and SS (Supplementary Methods). When the TF plots of an iEEG trial did not exhibit a power magnitude maximum meeting this threshold it was denoted as a null event. For trials with sufficient power magnitudes, the time and frequency coordinates  150 msec corresponding to the maximum power magnitude were considered as the candidate event epoch, but not the event itself. The candidate event epoch was subsequently processed using the topographical analysis\textsuperscript{21 }to define the event and when appropriate the power, duration, and frequency of the event. We did not seek to characterize multiple events in single trials, but rather only the event of largest spectral power. \par}\par

\vspace{\baselineskip}
{\fontsize{11pt}{13.2pt}\selectfont \textit{Classifying and characterizing ripple events in iEEG trials using topographical analysis}\par}\par

{\fontsize{11pt}{13.2pt}\selectfont To determine whether the candidate ripple event was a true ripple or a result of filter ringing from a sharply contoured spike or artifact\textsuperscript{21}, we applied a topographical analysis to the time-frequency (TF) plot\textsuperscript{2}. To develop an automatic software method for classifying ripples as true or false, we utilized the difference in the TF representation of sharp transients and true high frequency oscillations. TF plots of time series data exhibit an inherent topography defined by isopower contours. A true ripple is represented by a $``$blob$"$  of power within the ripple band (80-250 Hz)\textsuperscript{3}, and if contour lines are defined for a TF representation of the $``$blob$"$ , with the maximum and minimum frequencies confined to the ripple band, the contours will have closed loops. In contrast, a false ripple is represented by a $``$candle$"$  of power in the ripple band\textsuperscript{21} (Figure 1D), but importantly this $``$candle$"$  continues below the ripple band. Thus, when the contour lines are defined for the $``$candle$"$  within the ripple band, the contours will have open loops (Supplementary Methods). This same method was used to derive the mean spectral content, duration, and power of each ripple event (Supplementary Methods). Since some of the EEG amplifiers used in this study had an anti-aliasing filter at 125 Hz, we removed all the ripple events with a mean frequency greater than 120 Hz from the dataset. \par}\par

\vspace{\baselineskip}
{\fontsize{11pt}{13.2pt}\selectfont \textit{Identification of interictal discharges in the iEEG using a topographical analysis of time-frequency plots}\par}\par

{\fontsize{11pt}{13.2pt}\selectfont Both true and false ripple events were subsequently processed by a second stage algorithm designed to identify the presence of an interictal discharge, within 200 ms of the ripple event, on the basis of an analysis of TF plots resulting from wavelet convolution\textsuperscript{4}. This algorithm operated under the principle that interictal discharges produce a recognizable motif in the TF plot that is relatively independent of both the amplitude and slope of the iEEG during the discharge (Supplementary Methods). \par}\par

\vspace{\baselineskip}
{\fontsize{11pt}{13.2pt}\selectfont \textit{Repeated measures logistic regression models (LRM) for word encoding trials}\par}\par

\tab {\fontsize{11pt}{13.2pt}\selectfont As predictors in the statistical models, the absence of a ripple was coded ‘0’, ripple on spike (RonS) was coded ‘1’, spike was coded ‘2’, and ripple on oscillation (RonO) was coded ‘3’. The events recorded from individual electrodes were aggregated across the corresponding regions of interest. In the case of multiple events occurring within a single region, codes of higher number took precedence. If a given subject did not have coverage of a neuroanatomical region, then the data from that subject was missing for the region. The events of each word recall and proportions of word recall events in 12-word blocks were analyzed using the generalized linear mixed model (GLMM) model with the assumption of binomial distribution. The generalized estimating equation (GEE) estimation approach for population-average GLMM was used with the exchangeable working correlation matrix to account for correlations between rates of recall in repeated words within blocks, repeated blocks within experiment, and repeated experiments within the same subject. The GLMM model allows for variable number of observations per subject representing variable number of experiments and word blocks per patient. Separate models were fitted for each neuroanatomical region of interest. For each region, predictors of individual word recall included, the type of event and location of the corresponding electrode inside or outside the seizure-onset zone (SOZ), or whether the recording was made using a XLTEK EMU128 amplifier with a 131 Hz anti-aliasing filter (Natus, Pleasonton, CA). Predictors of proportions of word recall events in 12-word blocks included the type of events and the mean rates of RonS, Spike, and RonO per minute in the left temporal neocortex during the encoding, distractor and recall periods. If no events of interest occurred during the encoding or distractor or recall periods, then the corresponding predictors were equal to zero. The p-values from each set of models (including models for multiple regions of interest) were corrected for multiple comparisons using the adaptive Hochberg algorithm\textsuperscript{22}. The data were analyzed in R (Vienna, Austria) and SAS 9.4 (Cary, NC).\par}\par

\vspace{\baselineskip}
{\fontsize{11pt}{13.2pt}\selectfont \textit{Block design repeated measures logistic regression models }\par}\par

{\fontsize{11pt}{13.2pt}\selectfont The GEE estimation approach for population-average GLMM was used with the exchangeable working correlation matrix to account for correlations between rates of recall in repeated blocks within the same subject. The predictors of word recall included the mean rates of RonS, Spike, and RonO per minute in the left temporal neocortex during the encoding, distractor and recall periods. If no events of interest occurred during one of these\ periods, then the corresponding predictors were equal to zero.  Separate models were fitted using the mean rates for RonS, Spike, and RonO, because of linear dependency. The rates of word recall events in 12-word blocks were analyzed using the logistic regression model with assumption of the binomial distribution. The data analysis was performed in SAS (Cary, NC) and R (Vienna, Austria). Multiple testing correction was performed using the adaptive Hochberg algorithm in SAS\textsuperscript{22}.\par}\par

\vspace{\baselineskip}
{\fontsize{11pt}{13.2pt}\selectfont \textbf{Results }\par}\par

{\fontsize{11pt}{13.2pt}\selectfont Across 58,312 word trials of the free recall task (Figure 1A) the estimated overall word recall probability by session was 25.4  11.6$\%$  (s.e.m). We applied a detector that implemented a wavelet topographical algorithm\textsuperscript{20,23} to detect and quantify ripple and spike events in depth iEEG recordings from 7,630 unique locations (Figure 1B) in 107 patients (Supplemental Table 1). The algorithm specified whether an iEEG trial beginning 750 msec prior to word presentation and ending at word display offset contained a) neither a ripple nor spike, b) a ripple on spike (RonS), c) a sharply contoured epileptiform spike, d) a ripple not on a spike \textit{i.e.} a ripple on background oscillation (RonO)(Figure 1C,D). The detector could not quantify multiple events per trial. The detector identified only ripple events over a pre-determined power threshold corresponding to an amplitude threshold of >8 V. We selected a relatively high threshold to maximize the specificity of the detector for ripple events likely to be annotated by a clinical epileptogist\textsuperscript{15}. The detector was applied to iEEG recordings from all the experimental trials across all patients and at every non-excluded recording electrode site. We then calculated the sensitivity, precision, and inter-rater reliability as measured by the intraclass correlation coefficient of the detector for the different event types in mesial temporal and neocortical sites in 448 electrodes from 8 patients (Supplementary Methods, Supplemental Table 2). Since the human validation of the detector was not blinded, but rather performed by adding and deleting marked events, the moderately high performance of the detector (sensitivity $ \sim $ 70-80$\%$ , specificity $ \sim $ 70-80$\%$ , Supplementary Table 2) was inflated. \par}\par

{\fontsize{11pt}{13.2pt}\selectfont We examined the probability and rate of ripple and spike events by electrode (Figure 1C), and by the lobe in which they occurred. The event rates computed by lobe were preprocessed for use in the logistic regression models (LRMs, see methods). Thus, these rates do not precisely correspond with event rates present in the continuous iEEG recordings in each electrode recording site (Supplementary Table 3). Overall, ripple events occurred with the greatest rate and probability in temporal and limbic (\textit{i.e.} mesial temporal including the cingulate gyrus) regions bilaterally (Supplemental Table 3). At the single electrode level, this was also the case (Figure 1C). Since the bilateral temporal and limbic regions had superior electrode coverage and mostly higher rates of spike and ripple events, we focused our study on these regions to limit errors from limited spatial and temporal sampling due to infrequent events.\par}\par

{\fontsize{11pt}{13.2pt}\selectfont To determine whether ripples and spikes in the bilateral temporal and limbic regions during encoding modulate the probability of recall, we constructed four repeated measures logistic regression models (LRMs). Since prior work has demonstrated that spikes occurring outside the seizure onset zone (SOZ) correlate with disrupted memory encoding, while spikes within the SOZ do not\textsuperscript{24}, we included the SOZ as a covariate in the LRMs. In our study \textcolor[HTML]{1B2432}{the occurrence of RonS events was increased in the SOZ (}SOZ: 0.3$\%$  of trials, non-SOZ: 0.09$\%$ , t=2.13, d.f.=81, p=0.036, paired t-test\textcolor[HTML]{1B2432}{), but this was not the case for spikes or RonO events. }Using the LRMs we first asked if spikes or ripples within the SOZ correlate with disrupted encoding (Figure 2, Supplementary Table 4). The LRM for ripple and spike \textcolor[HTML]{1B2432}{events in the left temporal neocortex revealed that RonS events within the SOZ correlate with disrupted encoding (odds ratio [OR] of word recall 0.71, 95$\%$  confidence interval (CI): 0.59-0.85, n=158). This effect remained significant after correction for multiple testing (p\textsubscript{raw} =\textsuperscript{ }0.0001, p\textsubscript{adj }= 0.001, adaptive Hochberg method, n=24 tests). }\par}\par

{\fontsize{11pt}{13.2pt}\selectfont \textcolor[HTML]{1B2432}{We next utilized the four LRMs to ask if ripple and spike events in the non-SOZ correlate with disrupted encoding (Figure 3, Supplementary Table 4). Using the LRM for the left temporal neocortex we found that RonO events in the non-SOZ correlate with disrupted encoding (OR 0.55, 95$\%$  CI: 0.39-0.77, n=140 events, p\textsubscript{adj}=0.003). We also observed, using the LRM for ripple and spike events in the right limbic regions, that spikes in the right limbic non-SOZ correlate with disrupted encoding (OR 0.74, 95$\%$  CI: 0.60-0.92, n=524 events, p\textsubscript{adj}=0.036). }\par}\par

{\fontsize{11pt}{13.2pt}\selectfont \textcolor[HTML]{1B2432}{Since ripples significantly correlated with disrupted encoding in the left temporal neocortex, and the left temporal neocortex is likely to be involved in verbal episodic memory encoding\textsuperscript{24}, we next asked whether ripples and spikes correlate with disrupted encoding in specific sub-regions of the left temporal neocortex (Figure 4, Supplementary Table 5). We constructed LRMs of the superior, middle, and inferior left temporal gyrus, but did not use the SOZ as a covariate in these LRMs because of sample size limitations due to electrode coverage. Using the middle temporal gyrus model, we found that RonS (OR 0.53, 95$\%$  CI 0.34-0.82, n=125, p\textsubscript{adj}=0.023, adaptive Hochberg method, n=9 tests), and spikes (OR 0.61, 95$\%$  CI 0.53-0.34, n=743, p\textsubscript{adj}<0.001) correlate with disrupted encoding. Using the inferior temporal gyrus model, we found that spikes also correlate with disrupted encoding (OR 0.51, 95$\%$  CI 0.33-0.78, n=106, p\textsubscript{adj}=0.01). We also observed that RonO significantly correlated with disrupted encoding in the superior temporal gyrus (OR 0.60, 95$\%$  CI0.44-0.83, n=33, p\textsubscript{adj}=0.009, p\textsubscript{adj}=0.009). }\par}\par

{\fontsize{11pt}{13.2pt}\selectfont \textcolor[HTML]{1B2432}{Since we observed that ripples with a mean frequency < 120 Hz recorded from the left temporal neocortex correlate with disrupted encoding, we next asked if including ripple events with a higher frequency content would influence the effect, and if a 131 Hz anti-aliasing filter utilized for a portion of the recordings, showed any interaction with the effect. We found a trend showing that RonS and RonO events that occurred during encoding in the left temporal neocortex, irrespective of the location of the SOZ, corresponded with a decreased probability of recall (p\textsubscript{adj}<0.07). We also found no significant interaction between the use of an anti-aliasing filter and the effect of RonS and RonO events on recall probability (Supplementary Table 6). }\par}\par

\tab {\fontsize{11pt}{13.2pt}\selectfont We next sought to determine whether ripple and spike events that occur in the left temporal neocortex decrease the probability of recall only when they occur during the encoding epoch. We utilized a distinct LRM to correlate the mean rate of ripple events or spikes that occurred during either the encoding, distractor, or recall epochs in the left temporal neocortex with the number of items recalled per word encoding block (Figure 5, Supplemental Table 7). A distinct LRM was required because the distractor and recall epochs were not structured into trials. We could not use the SOZ as a covariate in this LRM because of sample size limitations due to the number of patients (N=42). We found that increases in the mean rate of RonS events in the left temporal neocortex during encoding, but not during the distractor, or recall, resulted in a decrease in the odds of recall (CI=0.92, 95$\%$  CI 0.9-0.95, p\textsubscript{adj}<0.001, step up Bonferroni method, n=18 tests). The same correlation was also seen for RonO events (OR=0.79, 95$\%$  CI 0.71-0.89, p\textsubscript{adj}<0.001, step up Bonferroni method, n=18 tests) (Figure 5, Supplementary Table 6,). In the case of spikes, a trend was evident during the encoding epoch (OR=0.94 95$\%$  CI 0.89-1.0, p\textsubscript{adj}=0.137), and a correlation was evident during the distractor epoch, but the effect size was small (OR=0.96, 95$\%$  CI 0.95-0.97, p\textsubscript{adj}<0.001). \par}\par

\vspace{\baselineskip}
{\fontsize{11pt}{13.2pt}\selectfont \textbf{Discussion }\par}\par

\tab {\fontsize{11pt}{13.2pt}\selectfont We demonstrate that during a list learning free recall task, when ripple events occur in the left temporal neocortex during word presentation the probability of successful recall is decreased. While other work has shown ripples promote memory consolidation\textsuperscript{11}, the criteria for distinguishing the ripples that promote memory from those that correlate with disrupted memory encoding are not yet clear. The neuroanatomical location wherein the ripple occurs appears to be important, as does the location of the seizure onset zone. Paradoxically, left temporal neocortex ripples on spikes (RonS) only correlated with disrupted encoding when they occurred at the site of seizure generation, whereas left temporal ripples on background EEG (RonO) only correlated with disrupted encoding when they occurred outside the site of seizure generation. Furthermore, in the left temporal neocortex, ripples that occurred during the distractor and recall epochs appeared to have no effect on memory performance. \par}\par

\tab {\fontsize{11pt}{13.2pt}\selectfont Our results suggest that, in patients with focal epilepsy, ripple events may serve both as a marker of memory processing\textsuperscript{2} and as a marker of memory interference. We did not explicitly distinguish the putative pathological ripples from other physiological higher-frequency oscillations in this study\textsuperscript{15}. Since we sought to identify only those events that would be acceptable for annotation during visual inspection of the iEEG by a clinical epileptologist, we utilized an automated ripple detector that identified only relatively high amplitude events\textsuperscript{20}. Both visual and automated ripple annotation have demonstrated that ripple rates are elevated in the SOZ during sleep and under anesthesia\textsuperscript{15}. We also found that, during behavior, RonS event probabilities were elevated in the SOZ. However, it is unlikely that our detector defined events that were exclusively pathological, since large amplitude ripple events detected in hippocampal iEEG recordings have also been found to positively correlate with memory consolidation performance\textsuperscript{11}. One difference between the ripple events defined in this study and the other studies examining the role of high-frequency activity in cognition\textsuperscript{12-14}, is that the events defined in this study occurred 2- to 3-fold less often due to the relatively high amplitude threshold of our detector. \par}\par

{\fontsize{11pt}{13.2pt}\selectfont In the left temporal neocortex, only RonS in the SOZ and RonO events outside the SOZ were associated with encoding disruption. However, the block design LRM (Figure 5), demonstrated that RonS and RonO events in the left temporal neocortex are associated with encoding disruption, irrespective of the location of the SOZ.  Ung et al.\textsuperscript{24} recently reported that spikes outside the SOZ in the left hemisphere during encoding resulted in a decrease in the probability of recall. Ung et al. theorized that this observation supports the concept of a ‘nociferous cortex’\textsuperscript{24} in which pathological activity interrupts normal electrophysiology and functioning in a region extending far outside of the primary epileptogenic zone. While Ung et al., did not measure ripple oscillations, our results support the ‘nociferous cortex’ concept, and highlight that activity associated with RonO events, correlate with disrupted cognition in the nociferous cortical territory. The mechanism by which ripple oscillations associate with a disruption of cognition is not yet established. A single prior study reported elevated ripple rates outside the SOZ in patients with impaired memory, while for patients with intact memory, this was not the case\textsuperscript{17}.\par}\par

{\fontsize{11pt}{13.2pt}\selectfont We also found that ripples and spikes in the left middle temporal gyrus (MTG) correlated with the most significant decrease in probability of successful recall, irrespective of the location of the SOZ. The left MTG is activated by tasks that probe semantic processing and memory\textsuperscript{25}, and lesion-mapping studies have shown that it is essential for word-level comprehension and retrieval\textsuperscript{26}. Perhaps the left MTG can be considered the most critical hub in the network of active and coordinated brain regions that mediate verbal episodic memory. Recently it was demonstrated that electrical stimulation of the left MTG during verbal episodic memory encoding can enhance the probability of recall\textsuperscript{27-29}.\par}\par

{\fontsize{11pt}{13.2pt}\selectfont The results of the LRMs did not demonstrate a significant correlation between ripples or spikes in mesial temporal structures and failed encoding, except in the case of spikes in the right mesial temporal and cingulate non-SOZ. However, the effect size in this region was relatively small. While some past studies have shown that spikes in the hippocampus correlate with disrupted encoding\textsuperscript{4-5}, other studies failed to replicate these findings\textsuperscript{6}. One possible explanation for this discrepancy is that if spatial granularity is too coarse in a study that utilizes pathologic events to map functional memory networks, the anatomy of that functional network may not be accurately characterized. In our study, we grouped all mesial temporal structures as well as the cingulate gyrus together in one category. We did not have the sample size required to examine the effects in entorhinal cortex and sub-regions of the hippocampus. Ripples and spikes in discrete mesial temporal regions and sub-regions may have diverging effects. Future investigations that utilize a larger number of patients, and accurate segmentation of mesial temporal structures prior to electrode localization, could hopefully resolve this controversy by demonstrating the effect of pathologic events in key structures in the mesial-temporal lobe. \par}\par

{\fontsize{11pt}{13.2pt}\selectfont The design of our study had several other shortcomings. In an effort to determine whether a ripple or spike would correlate with disrupted encoding, we deliberately analyzed the 750 msec preceding the word presentation, in addition to the word presentation. We included the iEEG recordings prior to the presentation because neuronal action potentials can be suppressed for 0.5-2 seconds following an epileptiform discharge\textsuperscript{30,31}. We did not relate events in the iEEG which occur during the 750 msec following word presentation to the recall of that word, despite the fact that encoding of that word may still be taking place. Another shortcoming was that we identified ripple and spike events in the entire encoding epoch, and did not explicitly examine the differential effects of events occurring at different stages or segments. For example, it is unclear if events in the occipital lobe would disrupt encoding only if the event occurs prior to or during the initial stages of word presentation\textsuperscript{32}. In addition, our design made it impossible to quantify the impact of multiple ripples or spikes that occur during a single trial, and the cognitive effects of the wave component of spike-wave discharges\textsuperscript{33}. \par}\par

{\fontsize{11pt}{13.2pt}\selectfont Another major shortcoming of our study was that the iEEG sampling rate was only 500 Hz thereby reducing the probability of detecting and quantifying ripples with a frequency greater than 120 Hz. We found that among the ripples detected and quantified in this study most of the events had frequencies < 120 Hz. This finding is in accord with a large number of studies which quantified the properties of ripples measured using macroelectrodes in humans at sampling rates of 2000 Hz\textsuperscript{12,16,20}. The limited sampling rate of this study eliminated the possibility of detecting and quantifying fast ripples. Also, fast ripples may have been mistakenly identified as ripples due to aliasing. Fast ripples selectively disrupt single neuron reactivation during SpW-R in the rodent hippocampus\textsuperscript{34}, and could also disrupt human episodic verbal memory and spatial memory. \par}\par

{\fontsize{11pt}{13.2pt}\selectfont In summary, both epileptiform spikes and ripples correlate with a failure of verbal episodic memory encoding in the left temporal neocortex and may contribute to cognitive disorders in temporal lobe epilepsy\textsuperscript{35}. Future studies should be directed at understanding the mechanistic differences that distinguish pathological ripples from physiological ripples\textsuperscript{36}, and high-frequency activity\textsuperscript{13.14,27}. One strategy may be to examine differences in single unit activity during the high-frequency activity and ripple events that coincide with successful and failed verbal memory encoding. \par}\par

\vspace{\baselineskip}
{\fontsize{11pt}{13.2pt}\selectfont \textbf{Data and software availability}\par}\par

{\fontsize{11pt}{13.2pt}\selectfont Analyzed data is available on a permanent Zenodo repository \href{https://www.zenodo.org/record/838790}{https://www.zenodo.org/record/838790$\#$ .WYStg9PyuWY}, $\#$ 2. Matlab, R, and SAS code used for data analysis is permanently available on Github \par}\textcolor[HTML]{353535}{https://github.com/shennanw/waldman\_RAM/}{\fontsize{11pt}{13.2pt}\selectfont .\par}\par

\vspace{\baselineskip}
{\fontsize{11pt}{13.2pt}\selectfont \textbf{Acknowledgements}\par}\par

{\fontsize{11pt}{13.2pt}\selectfont We\ thank Blackrock Microsystems for providing neural recording and stimulation equipment. This work was supported by the DARPA Restoring Active Memory (RAM) program (Cooperative Agreement N66001-14-2-4032).  The views, opinions, and/or findings contained in this material are those of the authors and should not be interpreted as representing the official views or policies of the Department of Defense or the U.S. Government. Dr. Weiss is supported by 1K23NS094633-01A1.\par}\par

 %%%%%%%%%%%%  Starting New Page here %%%%%%%%%%%%%%

\newpage

\vspace{\baselineskip}{\fontsize{11pt}{13.2pt}\selectfont \textbf{Conflicts of Interest}\par}\par

{\fontsize{11pt}{13.2pt}\selectfont \textcolor[HTML]{212121}{D.S.R. holds more than 5$\%$  equity interest in Nia Therapeutics, LLC ($``$Nia$"$ ) a brain stimulation device manufacturer}. S.A.W. and Z.J.W. both hold more than 5$\%$  equity interest in Fastwave L.L.C., a EEG software manufacturer. \par}\par

\vspace{\baselineskip}
{\fontsize{11pt}{13.2pt}\selectfont \textbf{Ethical Publication Statement}\par}\par

{\fontsize{11pt}{13.2pt}\selectfont The authors have read the journal’s position on issues involved in ethical publication and affirm their report is consistent with those guidelines.\par}\par

\vspace{\baselineskip}
{\fontsize{11pt}{13.2pt}\selectfont \textbf{REFERENCES}\par}\par

\vspace{\baselineskip}
\begin{enumerate}
	\item Doron G, von Heimendahl M, Schlattmann P, et al. Spiking irregularity and frequency modulate the behavioral report of single-neuron stimulation. \textit{Neuron} 2014;81:653-63.\par

	\item Buzsáki G. Hippocampal sharp wave-ripple: A cognitive biomarker for episodic memory and planning. \textit{Hippocampus} 2015;25:1073-188. \par

	\item Aarts JH, Binnie CD, Smit AM, et al. Selective cognitive impairment during focal and generalized epileptiform EEG activity. \textit{Brain} 1984;107:293-308.\par

	\item Krauss GL, Summerfield M, Brandt J, et al. Mesial temporal spikes interfere with working memory. \textit{Neurology} 1997;49:975-80.\par

	\item Kleen JK, Scott RC, Holmes GL, et al. Hippocampal interictal epileptiform activity disrupts cognition in humans. \textit{Neurology} 2013;81:18-24.\par

	\item Horak PC, Meisenhelter S, Song Y, et al. Interictal epileptiform discharges impair word recall in multiple brain areas. \textit{Epilepsia} 2017;58:373-380.\par

	\item O'Neill J, Senior T, Csicsvari J. Place-selective firing of CA1 pyramidal cells during sharp wave/ripple network patterns in exploratory behavior. \textit{Neuron} 2006;49:143-55.\par

	\item Jadhav SP, Kemere C, German PW. Awake hippocampal sharp-wave ripples support spatial memory. \textit{Science} 2012;336:1454-8.\par

	\item Carr MF, Jadhav SP, Frank LM. Hippocampal replay in the awake state: a potential substrate for memory consolidation and retrieval. \textit{Nat Neurosci} 2011;14:147-53.\par

	\item Nakashiba T, Buhl DL, McHugh TJ, et al. Hippocampal CA3 output is crucial for ripple-associated reactivation and consolidation of memory. \textit{Neuron} 2009;62:781-7.\par

	\item Axmacher N, Elger CE, Fell J. Ripples in the medial temporal lobe are relevant for human memory consolidation. \textit{Brain} 2008;131:1806-17. \par

	\item Matsumoto A, Brinkmann BH, Matthew Stead S, et al. Pathological and physiological high-frequency oscillations in focal human epilepsy. \textit{J Neurophysiol} 2013;110:1958-64.\par

	\item Kucewicz MT, Cimbalnik J, Matsumoto JY, et al. High frequency oscillations are associated with cognitive processing in human recognition memory. \textit{Brain} 2014;137:2231-44.\par

	\item Kucewicz MT, Berry BM, Kremen V, et al. Dissecting gamma frequency activity during human memory processing. \textit{Brain} 2017;140:1337-1350.\par

	\item Frauscher B, Bartolomei F, Kobayashi K, et al. High-frequency oscillations: The state of clinical research. \textit{Epilepsia} 2017;58:1316-1329.\par

	\item Weiss SA, Orosz I, Salamon N, et al. Ripples on spikes show increased phase-amplitude coupling in mesial temporal lobe epilepsy seizure-onset zones. \textit{Epilepsia} 2016;57:1916-1930.\par

	\item Jacobs J, Banks S, Zelmann R, et al. Spontaneous ripples in the hippocampus correlate with epileptogenicity and not memory function in patients with refractory epilepsy\textit{. Epilepsy Behav} 2016;62:258-66.\par

	\item Jacobs J, Miller J, Lee SA, et al. Direct Electrical Stimulation of the Human Entorhinal Region and Hippocampus Impairs Memory. \textit{Neuron} 2016;92:983-990.\par

	\item Dvorak D, Fenton AA. Toward a proper estimation of phase–amplitude coupling in neural oscillations. \textit{J Neurosci Methods} 2014;225:42–56. \par

	\item Waldman Z, Shimamoto S, Song I, et al. A Method for the Topographical Identification and Quantification of High Frequency Oscillations in Intracranial Electroencephalography Recordings. \textit{Clinical Neurophysiology} (in press 2017)\par

	\item Bénar CG, Chauvière L, Bartolomei F, et al. Pitfalls of high-pass filtering for detecting epileptic oscillations: a technical note on "false" ripples. \textit{Clin Neurophysiol} 2010; 121:301-10.\par

	\item Hochberg, Y. and Benjamini, Y. (1990), $``$More Powerful Procedures for Multiple Significance Testing,$"$  Statistics in Medicine, 9, 811–818.\par

	\item Shimamoto S, Waldman Z, Orosz I, et al. Utilization of independent component analysis for accurate pathological ripple detection in intracranial EEG recordings recorded extra- and intra-operatively. \textit{Clinical Neurophysiology} \textit{(}in press 2017)\par

	\item Ung H, Cazares C, Nanivadekar A, et al. Interictal epileptiform activity outside the seizure onset zone impacts cognition. \textit{Brain} 2017;140:2157-2168.\par

	\item Wei T, Liang X, He Y, et al. Predicting conceptual processing capacity from spontaneous neuronal activity of the left middle temporal gyrus. \textit{J Neurosci} 2012;32:481-9.\par

	\item Baldo JV, Arévalo A, Patterson JP, et al. Grey and white matter correlates of picture naming: evidence from a voxel-based lesion analysis of the Boston Naming Test. \textit{Cortex} 2013; 49:658-67. \par

	\item Kucewicz MT, Berry BM, Miller LR, Khadjevand F, Ezzyat Y, Stein JM, Kremen V, Brinkmann BH, Wanda P, Sperling MR, Gorniak R, Davis KA, Jobst BC, Gross RE, Lega B, Van Gompel J, Stead SM, Rizzuto DS, Kahana MJ, Worrell GA. Evidence for verbal memory enhancement with electrical brain stimulation in the lateral temporal cortex. Brain. 2018 Jan 8. doi: 10.1093/brain/awx373.\par

	\item Kucewicz MT, Berry BM, Kremen V, Miller LR, Khadjevand F, Ezzyat Y, Stein JM, Wanda P, Sperling MR, Gorniak R, Davis KA, Jobst BC, Gross RE, Lega B, Stead SM, Rizzuto DS, Kahana MJ, Worrell GA. Electrical Stimulation Modulates High $ \gamma $  Activity and Human Memory Performance. eNeuro. 2018 Feb 2;5(1). pii: ENEURO.0369-17.2018.\par

	\item Ezzyat Y, Wanda PA, Levy DF, Kadel A, Aka A, Pedisich I, Sperling MR, Sharan AD, Lega BC, Burks A, Gross RE, Inman CS, Jobst BC, Gorenstein MA, Davis KA, Worrell GA, Kucewicz MT, Stein JM, Gorniak R, Das SR, Rizzuto DS, Kahana MJ. Closed-loop stimulation of temporal cortex rescues functional networks and improves memory. Nat Commun. 2018 Feb 6;9(1):365.\par

	\item Keller CJ, Truccolo W, Gale JT, et al. Heterogeneous neuronal firing patterns during interictal epileptiform discharges in the human cortex. \textit{Brain} 2010;133:1668-81.\par

	\item Alarcón G, Martinez J, Kerai SV, et al. In vivo neuronal firing patterns during human epileptiform discharges replicated by electrical stimulation. \textit{Clin Neurophysiol} 2012;123:1736-44.\par

	\item Shewmon DA, Erwin RJ. The effect of focal interictal spikes on perception and reaction time. II. Neuroanatomic specificity. \textit{Electroencephalogr Clin Neurophysiol} 1988;69:338-52.\par

	\item Shewmon DA, Erwin RJ. Focal spike-induced cerebral dysfunction is related to the after-coming slow wave. \textit{Ann Neurol} 1988;23:131-7.\par

	\item Valero M, Averkin RG, Fernandez-Lamo I, et al. Mechanisms for Selective Single-Cell Reactivation during Offline Sharp-Wave Ripples and Their Distortion by Fast Ripples. \textit{Neuron} 2017;94:1234-1247.e7.\par

	\item Bell B, Lin JJ, Seidenberg M, Hermann B. The neurobiology of cognitive disorders in temporal lobe epilepsy. Nat Rev \textit{Neurol} 2011;7:154–164.\par

	\item Le Van Quyen M, Bragin A, Staba R,et al. Cell type-specific firing during ripple oscillations in the hippocampal formation of humans. \textit{J Neurosci} 2008;28:6104-10.
\end{enumerate}\par

\vspace{\baselineskip}

\vspace{\baselineskip}

\vspace{\baselineskip}

\vspace{\baselineskip}

\vspace{\baselineskip}

\vspace{\baselineskip}

\vspace{\baselineskip}

\vspace{\baselineskip}

\vspace{\baselineskip}

\vspace{\baselineskip}

\vspace{\baselineskip}

\vspace{\baselineskip}

\vspace{\baselineskip}

\vspace{\baselineskip}

\vspace{\baselineskip}

\vspace{\baselineskip}

\vspace{\baselineskip}

\vspace{\baselineskip}

\vspace{\baselineskip}

\vspace{\baselineskip}

\vspace{\baselineskip}

\vspace{\baselineskip}

\vspace{\baselineskip}

\vspace{\baselineskip}

\vspace{\baselineskip}

\vspace{\baselineskip}

\vspace{\baselineskip}

\vspace{\baselineskip}

\vspace{\baselineskip}

\vspace{\baselineskip}

\vspace{\baselineskip}

\vspace{\baselineskip}

\vspace{\baselineskip}
\textbf{\uline{Figures}}\par

\vspace{\baselineskip}

\vspace{\baselineskip}
\textbf{\uline{Figure 1}: Mapping ripple and spike events in a cohort of 107 medically refractory epilepsy patients performing a verbal episodic memory task.} \textbf{A. }Illustration of the verbal episodic memory task including the encoding, distractor, and recall epochs. \textbf{B. }Illustrative brain rendering demonstrating the number of electrode contacts in each lobe across the entire patient cohort. \textbf{C.} Superimposition of the location of the intracranial depth EEG recording contacts in the cohort of 107 patients in a glass brain. The color of each electrode indicates the probability of a word presentation trial in which a ripple on spike (top), ripple on oscillation (middle), or spike (bottom) occurs\ in the iEEG.  This event probability is proportionate to event rate irrespective of the task. \textbf{D.} Example time-frequency spectrograms and corresponding iEEG traces for each of the event types. Panel D1 illustrates a distinct $``$blob$"$ , representing the ripple, with a spectral content greater than a simultaneous $``$candle$"$ , representing the spike. Panel D2 illustrates a distinct $``$blob$"$ , representing the ripple. Panel D3 illustrates a distinct $``$candle$"$  representing the spike. The distinction between the $``$blob$"$  in D2 and $``$candle$"$  in D3 is indicated by respective white arrows. \par

\vspace{\baselineskip}

 %%%%%%%%%%%%  Starting New Page here %%%%%%%%%%%%%%

\newpage

\vspace{\baselineskip}
\vspace{\baselineskip}
\setstretch{1.0}
\textbf{\uline{Figure 2:}}{\fontsize{10pt}{12.0pt}\selectfont  \par}\textbf{Ripples on spikes in the left temporal neocortex seizure onset zone during word encoding decrease the odds of successful word recall.} Bar graph of the odds ratio and 95$\%$  confidence interval of successful word recall given the occurrence of ripple and spike events in the seizure onset zone during verbal encoding. When ripple and spike events occurred in the seizure onset zone during encoding, only ripple on spike events in the left temporal neocortex resulted in a significant decrease in the odds of word recall after a correction for multiple testing (adaptive Hochberg method, $\ast$ $\ast$ p<0.01, n=24 tests, N=number of patients, 10-50 blocks of 12 words per patient).\par

\vspace{\baselineskip}

\vspace{\baselineskip}

 %%%%%%%%%%%%  Starting New Page here %%%%%%%%%%%%%%

\newpage

\vspace{\baselineskip}\textbf{\uline{Figure 3:}}{\fontsize{10pt}{12.0pt}\selectfont  \par}\textbf{Ripples on oscillations in the left temporal neocortex non-seizure onset zone during word encoding decrease the odds of successful word recall.} Bar graph of the odds ratio and 95$\%$  confidence interval of successful word recall given the occurrence of ripple and spike events in the non-seizure onset zone during verbal encoding. When ripple and spike events occurred in the non-seizure onset zone during encoding, ripple on oscillation events in the left temporal neocortex, and spikes in the right mesial-temporal and cingulate regions resulted in a significant decrease in the odds of word recall after correcting for multiple testing (adaptive Hochberg method, $\ast$ $\ast$ p<0.01, $\ast$ p<0.05, n=24 tests, N=number of patients, 10-50 blocks of 12 words per patient).\par

\vspace{\baselineskip}

 %%%%%%%%%%%%  Starting New Page here %%%%%%%%%%%%%%

\newpage

\vspace{\baselineskip}\textbf{\uline{Figure 4:}} \textbf{Spikes in the left middle temporal gyrus during encoding correlate most significantly with a decrease in the probability of correct recall.} Bar graph of the odds ratio and 95$\%$  confidence interval of successful word recall given the occurrence of ripple and spike events during verbal encoding, irrespective of the seizure onset zone. When ripple and spike events occurred in the left temporal neocortex during encoding, ripple and spike events in the left middle temporal gyrus, and spikes in the left inferior temporal gyrus resulted in a significant decrease in the odds of word recall after correcting for multiple comparisons (adaptive Hochberg method, $\ast$ $\ast$ p<0.01, $\ast$ p<0.05, n=6 tests, N=number of patients, 10-50 blocks of 12 words per patient).\par

\vspace{\baselineskip}

\vspace{\baselineskip}

 %%%%%%%%%%%%  Starting New Page here %%%%%%%%%%%%%%

\newpage

\vspace{\baselineskip}\textbf{\uline{Figure 5:} Ripple events significantly reduce the odds of correct word recall when the events occur during the word encoding epoch, but not the distractor or recall epochs. }Bar graph of the change in the odds ratio of word recall predicted by an increase in the rate of ripple and spike events in the left temporal neocortex during the encoding, distractor, and recall epochs irrespective of the location of the seizure-onset zone adjusted for multiple testing (adaptive Hochberg method, $\ast$ p<0.001, n=18 tests, N=42, 10-50 blocks of 12 words per patient). Odds ratio shown correspond to one event per minute increase in rates.\par

\vspace{\baselineskip}
 \par

\printbibliography
\end{document}